\documentclass[10pt]{iopart}

\usepackage{iopams}
\usepackage{graphicx}
\begin{document}

\title{Thin film deposition with time varying temperature}

\author{T. A. de Assis}
\address{Instituto de F\'{\i}sica, Universidade Federal da Bahia,
   Campus Universit\'{a}rio da Federa\c c\~ao,
   Rua Bar\~{a}o de Jeremoabo s/n,
40170-115, Salvador, BA, Brazil}
\ead{thiagoaa@ufba.br}

\author{F. D. A. Aar\~ao Reis}
\address{Instituto de F\'\i sica, Universidade Federal Fluminense, Avenida Litor\^anea s/n,
24210-340 Niter\'oi RJ, Brazil}
\ead{reis@if.uff.br}

\begin{abstract}
We study the effects of time-dependent substrate/film temperature in the deposition of
a mesoscopically thick film using a statistical model that accounts for diffusion of
adatoms without lateral neighbors whose coefficients depend on an activation energy and
temperature. Dynamic scaling with fixed temperature is extended to predict conditions
in which the temperature variation significantly affects surface roughness scaling.
It agrees with computer simulation results for deposition of up to ${10}^4$ atomic layers and
maximal temperature changes of $30 K$, near or below the room temperature.
If the temperature decreases during the growth, the global roughness may have a rapid growth,
with effective exponents larger than $1/2$ due to the time-decreasing adatom mobility.
The local roughness in small box size shows typical evidence of anomalous scaling,
with anomaly exponents depending on the particular form of temperature decrease.
If the temperature increases during the growth, a non-monotonic evolution of the global roughness
may be observed, which is explained by
the competition of kinetic roughening and the smoothing effect of increasing diffusion lengths.
The extension of the theoretical approach to film deposition with other activation energy barriers
shows that similar conditions on temperature variation may lead to the same morphological features.
Equivalent results may also be observed by controlling the deposition flux.

\end{abstract}

\pacs{68.55.-a, 68.35.Ct, 81.15.Aa , 05.40.-a}

\maketitle

\section{Introduction}
\label{intro}

The morphological properties of a growing film by vapor techniques are mainly
determined by the balance between substrate/film temperature and pressure.
High temperature helps the system
to attain equilibrium states, for instance favoring formation of smooth surfaces
in homoepitaxy, while the increase of pressure leads to faster adsorption and drives the
system far from the equilibrium conditions \cite{ohring}. For these reasons,
many statistical models of thin film growth represent the competition of the external
flux of atoms or molecules and temperature-dependent surface processes, such as diffusion,
aggregation, and reactions \cite{pimpinelli,barabasi,etb}.

The main quantity to characterize surface morphology is the roughness, which measures thickness
fluctuations along the film surface.
It may be measured for the whole surface (global roughness) or inside a box that slides
on that surface (local roughness). The scaling properties with time
and size are described by dynamic relations \cite{barabasi}, which are called
normal when local and global fluctuations scale with the same exponents \cite{fv} and anomalous
when local and global exponents are different \cite{huo,schwarzacher,lopez,lopezPRE1996,auger}.

A question of experimental relevance is the effect of changing physico-chemical conditions
during film growth. For example, a recent experimental work on electrodeposition of Prussian
Blue films interpreted the growth as a process dominated by surface diffusion in which
the time-increasing adsorption rate is responsible for an anomaly in the dynamic scaling relation
\cite{PB}. The problem of spontaneous fluid imbibition in a porous medium is another
system that shows anomalous scaling (AS) and is modeled as interface growth with time-dependent
couplings \cite{alava2004}.

This scenario motivated theoretical works on growth models where the microscopic rules of
aggregation change in time. Ref. \protect\cite{chou} showed that a sudden change in the
parameters of the Edwards-Wilkinson (EW) growth equation \cite{ew} may be responsible
for nontrivial effects in roughness scaling, such as power-law relaxation to steady states.
The AS in stochastic growth equations with time-dependent couplings was discussed in
Ref. \protect\cite{pradas} and illustrated for
the EW equation with time-dependent surface tension, showing good agreement with
numerical results for models of spontaneous imbibition \cite{dube,pradas2006}.
In lattice models of film growth, AS was recently shown in competitive models with
time-varying probabilities \cite{anomcompet}.

On the other hand, to our knowledge no work have already discussed the effects of time-varying
temperature in thin film deposition.
For this reason, this paper is devoted to study a model of deposition
and diffusion of adsorbed species in which the diffusion coefficients are affected by a time-dependent
substrate/film temperature.
For simplicity, the model assumes that only adatoms in terraces can move,
while atoms with lateral neighbors are permanently aggregated.
A combination of a theoretical approach and simulation results shows several nontrivial effects on surface
roughness scaling that may help to understand experimental results. In cases of decreasing temperature,
there is evidence of AS for film thicknesses typical of real mesoscopically thick films. Moreover,
the global roughness has a nontrivial evolution which does not allow a reliable calculation of a growth
exponent. In cases of increasing temperature,
a non-monotonic evolution of the global roughness may be observed, possibly with a saturation for a
wide range of film thickness. The main advance from previous works is to show those features in
feasible conditions of film growth, since the model is an approximate description of
low-temperature physical vapor deposition. Possible extensions to deposition models with several
activation energy barriers suggest that the same features may be observed in similar conditions.

The rest of this paper is organized as follows. In Sec. \ref{definitions}, we will define the model, review basic
concepts of dynamic scaling of surface roughness, and discuss properties of the model for fixed
temperature. In Sec. \ref{theoretical}, a theoretical approach for time-dependent temperature is
developed to predict conditions in which the roughness scaling is affected or not.
In Sec. \ref{decreasingT}, we will discuss results for some conditions of decreasing temperature.
In Sec. \ref{increasingT}, we will discuss results for some conditions of increasing temperature.
In Sec. \ref{extensions}, the extension of our approach to other models is discussed.
In Sec. \ref{conclusion}, we summarize our results and present our conclusions.

\section{Basic definitions and concepts}
\label{definitions}

\subsection{Model definition and simulations}
\label{model}

The model studied in this work was introduced in Ref. \protect\cite{CDLM}.
It is a solid-on-solid type model (no overhang in the film surface) and the deposit has
a simple cubic lattice structure. The substrate is flat, at the $xy$ plane, with linear size $L$
($L\times L$ columns), with the lattice parameter being the length unit. The column height
is the $z$ coordinate of the topmost adatom.
There is an external flux of $F$ atoms per site per unit time, measured in number of monolayers
(ML) per second.
An incident atom is adsorbed upon landing above a previously deposited atom or a substrate site.
All adsorbed atoms with no lateral and no upper neighbor diffuse with coefficient $D$
(number of steps per unit time). If an adatom has a lateral or an
upper neighbor, then it is permanently aggregated at that position.
Fig. 1 illustrates the possible steps of some mobile atoms at the film surface.

\begin{figure}
\begin{center}
\includegraphics [width=12.5cm] {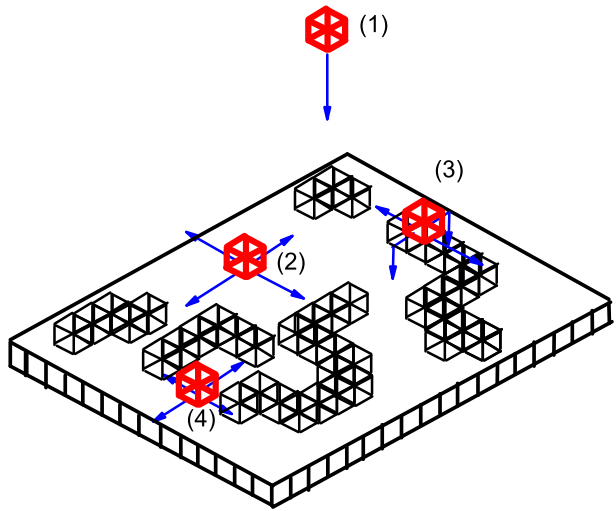}
\caption{Illustration of the model: deposition of an atom (1);
adatom diffusion in terraces (2), with possibility of downhill movement and
irreversible aggregation (3), and with possibility of irreversible aggregation at a step edge (4).}
\label{fig1}
\end{center}
\end{figure}

The diffusion coefficient $D$ depends on temperature as
\begin{equation}
D=\nu_0\exp{\left( -E/k_BT\right)} ,
\label{defD}
\end{equation}
where $\nu_0$ is a jump frequency
and $E$ is the activation energy in a flat surface, which typically amounts to tenths of $eV$.
The parameter
\begin{equation}
R\equiv D/F
\label{defR}
\end{equation}
quantitatively represents the
interplay between temperature and pressure and determines the film properties.
The other relevant parameter is the film thickness $t$, proportional to the deposition time.

In the submonolayer growth regime, this model corresponds to irreversible island growth
with critical nucleus of size $i=1$, which was already applied to several
systems \cite{etb}. The irreversible lateral aggregation is a reasonable assumption
for low temperature or large atomic flux, although it does not respect
detailed balance conditions. This model also has the advantage of reducing simulation times
compared with models where all atoms are mobile. The aggregation condition is similar to the Das Sarma
and Tamborenea (DT) model of molecular beam epitaxy \cite{dt}, but the main difference is
that the DT model and its extensions restrict the adatom diffusion to finite distances.
Moreover, the model does not consider Ehrlich-Schwebel (ES) barriers for downhill movement at terrace edges
\cite{es} and does not allow uphill movement because adatoms with lateral neighbors are immobile.
Indeed, our aim is to consider a model with a minimal
set of relevant parameters, in order to search for features
that are intrinsically related to the time-varying temperature.

Simulation results presented here were obtained in lattices with $L=512$. Some results were also
obtained in $L=256$ and $L=1024$ in order to confirm that finite-size effects are negligible
in the simulated time range. Fixed deposition parameters are
$\nu_0/F={10}^{14}$ and the activation energy $E=0.6 eV$.
These values are reasonable compared to experimental data for several metals and semiconductors,
although the direct applicability is limited due to the differences in lattice structure
and aggregation assumptions \cite{evansSS1993,etb}.
Choosing a fixed value of $E$ and changing $T$ does not restrict our conclusions because
the physically important parameter is $R$, which is a function of $E/T$.
In all cases, the temperature is uniform across the film and may vary in time.
Some results for fixed temperature are also obtained for comparison.

We will consider temperature values in which $R$ ranges between ${10}^1$ and ${10}^5$.
These values are small if compared to works on submonolayer growth, which frequently
consider $R$ up to ${10}^9$ \cite{etb}. However, they are suitable for low temperature deposition,
which is a necessary condition for the applicability of a model with irreversible attachment
to steps (in other words, even with a small binding energy to a lateral neighbor, the relative
probability of an attached atom to move will be small).
For $R={10}^5$, results of Ref. \protect\cite{CDLM} show that adatoms execute
an average of $28$ random steps before aggregation, thus diffusion lengths in
terraces are much smaller than the lateral size of the films and finite-size effects are negligible.

During the growth of a sample, the adsorption of each new atom takes place in a
time interval $1/\left( FL^2\right)$. After this process,
$R/L^2$ steps of randomly chosen free atoms are performed. Since $R/L^2$ is usually not integer,
and it may be small for large $L$, we keep its fractional part
for determining the number of steps after the next adsorption events.
If the temperature changes during the film growth, then the value of $R$ is updated after the
deposition of each complete layer ($L^2$ atoms), which corresponds to a unit time interval.

\subsection{Dynamic scaling of surface roughness}
\label{scaling}

The global roughness (or interface width) is defined as
\begin{equation}
W\left( L,t\right) \equiv { \left< {\overline{ \left( h - \overline{h}\right)} }^2
\right> }^{1/2} ,
\label{defw}
\end{equation}
where the overbars represent spatial averages and the angular brackets represent
configurational averages over various samples.
For short deposition times (small thicknesses), effects of the finite substrate size $L$ are
negligible and $W$ scales as
\begin{equation}
W\sim t^{\beta} ,
\label{defbeta}
\end{equation}
where $\beta$ is called the growth exponent.

For comparison with experimental data, the local roughness $w\left( r,t\right)$ is more useful.
It is also defined as a root mean square height fluctuation, but the spatial average is
limited to a box of size $r$, with $r<L$, and the configurational average is performed by
gliding this box over the film surface and over various samples.
In the regime of thin film growth (negligible effects of substrate size), it follows the
Family-Vicsek (FV) dynamic scaling relation \cite{fv}
\begin{equation}
w\left( r,t\right) \approx A r^{\alpha_l} g{\left( \frac{Bt}{r^{z}}\right)} .
\label{fvlocal}
\end{equation}
Here, $\alpha_l$ is the local roughness exponent (sometimes called Hurst exponent),
$z$ is the dynamic exponent, $B$ is a constant,
and $g$ is a scaling function such that $g(x)\sim 1$ for $x\gg 1$ (small box size) and
$g(x)\sim x^{-\alpha_l/z}$ for $x\ll 1$ (large box size, where the local roughness coincides with the
global one).

In systems with normal scaling, the amplitude $A$ in Eq. (\ref{fvlocal}) is a constant
(time-independent), thus $\beta=\alpha_l/z$ and the global roughness exponent $\alpha$
equals the local one. A scaling analysis of $w\left( r,t\right)$ of several
growth models in one- and two-dimensional substrates is presented in  Ref. \protect\cite{chamereis}.
In systems with AS, the amplitude $A$ scales as
\begin{equation}
A \sim t^\kappa ,
\label{defkappa}
\end{equation}
with $\kappa$ characterizing the degree of anomaly \cite{lopez}.
A FV relation can also be defined for the global roughness and involves the exponent $\alpha$,
but it is useful only if very long times / large thicknesses are attained.

\subsection{Previous works}
\label{previous}

Here we restrict the discussion to results on two-dimensional substrates (growth in $2+1$ dimensions),
which is the subject of the present work.

Several models with diffusion of all adatoms and energy barriers proportional to the number of
neighbors were previously studied in fixed temperature.
Some authors proposed that they had AS \cite{lanczycki} or a logarithmic scaling
of amplitude $A$ in the FV relation \cite{kotrla}.
Wilby et al \cite{wilby} showed that the
growth exponent $\beta$ was close to value of the universality class of the fourth order nonlinear
growth equation of Villain, Lai and Das Sarma (VLDS) \cite{laidassarma,villain}.
Recent renormalization studies explained the long crossover to VLDS scaling
\cite{hasel2007,haselPRE2008}.

The model presented in Sec. \ref{model} was already studied numerically, showing a FV relation for the
global roughness that includes the $R$-dependence as \cite{CDLM}
\begin{equation}
W = \frac{L^\alpha}{R^x} f{\left( R^yt/L^z\right) },
\label{fvcdlm}
\end{equation}
with $\alpha\approx 2/3$, $z\approx 10/3$, $x\approx 0.5$, and $y\approx 1$. The exponents $\alpha$
and $z$ agree with those of the VLDS class \cite{barabasi,janssen,crsosreis}.
The values of exponents $x$ and $y$ were explained by scaling arguments \cite{CDLM}.

For $R\geq {10}^5$, the surface roughness of this model is very small, even with very large thicknesses
(${10}^4$ or ${10}^5$ ML) \cite{CDLM}. This is not a regime of dynamic scaling, which is an additional
reason for our simulations to be performed with smaller values of $R$.
Full diffusion models with values of $R$ of the same order may show
larger roughness, particularly if ES barriers are included \cite{lealJSM2011}, but
with an extended set of parameters.

\subsection{Scaling of local roughness in constant temperature}
\label{local}

A recent work on numerical integration of the VLDS equation showed evidence of AS
in $2+1$ dimensions \cite{xia}, which raises the question on a possible AS in our lattice model.
This feature was not investigated in Ref. \protect\cite{CDLM}.

We performed simulations of the model with $10\leq R\leq 1000$ for fixed $R$ (fixed temperature).
Fig. 2 shows the local roughness as a function of box size for three values of that parameter.
For $R=10$, there is a split in the curves for small $r$, which suggests AS. However, that split
tends to disappear as the thickness increases. For instance, from $t=4000 ML$ to $t=8000 ML$, the
change in the local roughness for $r=5$ is only $7\%$.
For $R=100$ and $R=1000$, the split of the curves is almost negligible for $t>{10}^3$.
For those reasons, we understand that there is no asymptotic AS in our lattice model with fixed $R$,
and an apparent anomaly appears only for small $R$ and for small thicknesses due to some type
of scaling correction.

\begin{figure}
\begin{center}
\includegraphics [width=7.5cm] {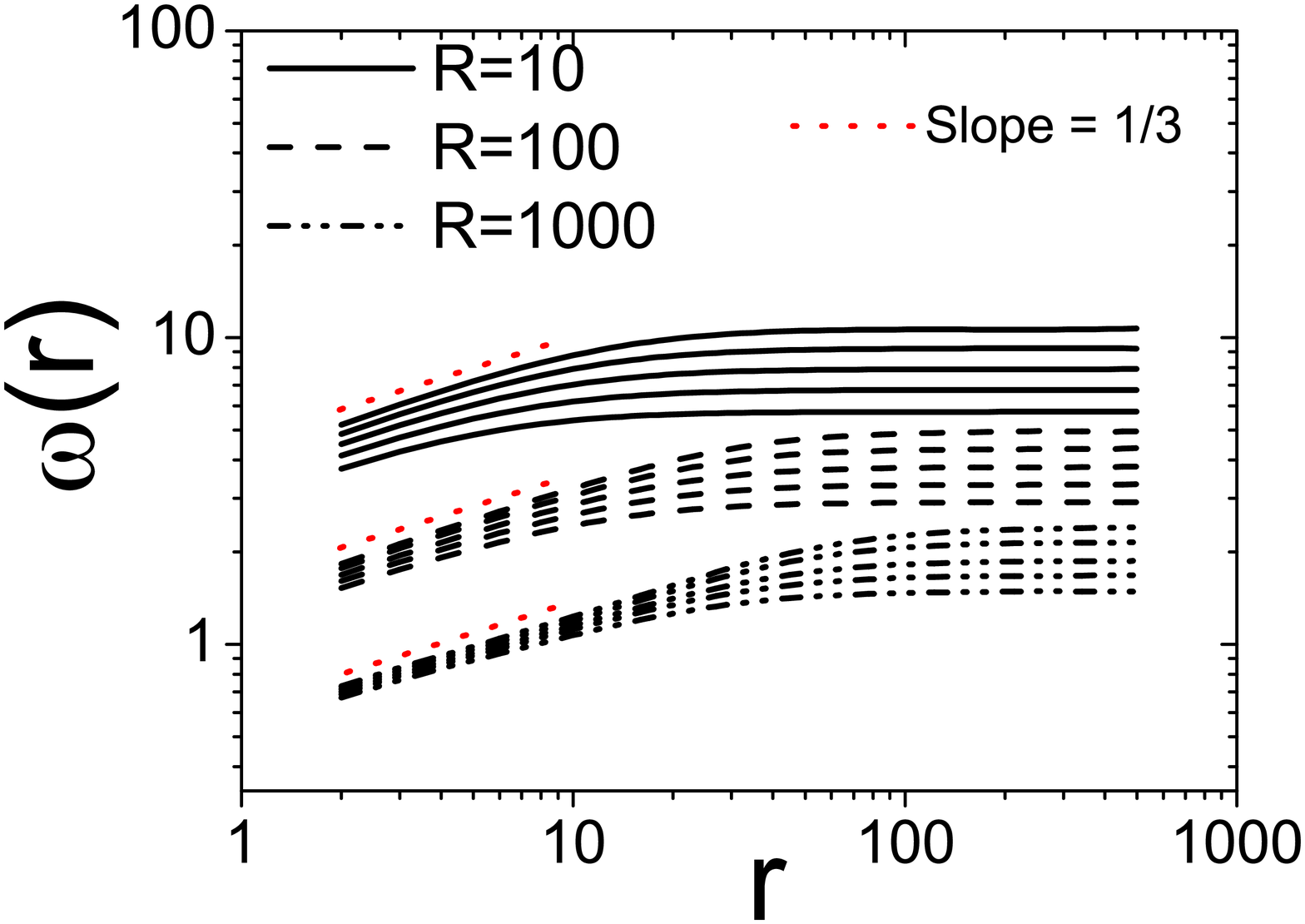}
\caption{Local roughness as a function of box size for R=10 (full line), R=100 (dashed line) and R=1000 (dashed dotted dotted line)
at times $t=500$ML, $1000$ML, $2000$ML, $4000$ML and $8000$ML, from bottom to top. The dotted line indicates the slope value of 1/3.}
\label{fig2}
\end{center}
\end{figure}
Fig. 2 also shows that the local roughness scales with
$\alpha_l\approx 1/3$ for small $r$ [Eq. (\ref{fvlocal})]. This exponent is significantly below the VLDS value
$\alpha_l=\alpha\approx 2/3$ \cite{crsosreis}. This discrepancy it not a failure of theoretical
predictions, but a consequence of scaling corrections also observed in other growth models \cite{chamereis}.

\section{Theoretical approach to temperature varying conditions}
\label{theoretical}

In competitive models involving two different aggregation dynamics, a FV relation similar to
Eq. (\ref{fvcdlm}) is obtained, with the probability $p$ replacing the ratio $R$ and exponents
$x$ and $y$ depending on the competing components \cite{albanordcor,rdcor}.
When $p$ slowly changes in time, the surface roughness evolution is predicted by
substituting the time-dependent form of that parameter in the corresponding FV relation
\cite{anomcompet}. This is supported by simulation results of several competitive models \cite{anomcompet}.

Relation (\ref{fvcdlm}) in the growth regime (where finite-size effects are negligible) gives
\begin{equation}
W = B \frac{t^{\beta}}{R^{\Delta}} \qquad ,\qquad  \beta=0.2 \qquad , \qquad \Delta=0.3 ,
\label{growthcdlm}
\end{equation}
where $B$ is a constant and the exponents $\beta$ and $\Delta$
were obtained from the values of $\alpha$, $z$, $x$, and $y$.
The data for constant $R$ gives $B\approx 3.3$ \cite{CDLM}.

In order to understand the conditions in which a temperature change has a significant effect in
the roughness evolution, we consider the rate of variation of the surface roughness:
\begin{equation}
\frac{dW}{dt} = B \frac{t^{\beta-1}}{R^{\Delta}} \left[ \beta - \Delta s\left( t\right) \right] \qquad ,\qquad
s(t)\equiv \frac{E}{k_BT}\frac{d\ln{T}}{d\ln{t}} .
\label{dwdt}
\end{equation}
The extension of Eq. (\ref{growthcdlm}) to time varying $R$ is a reasonable assumption only if
the roughness change is of the same order or smaller than that of the fixed temperature case.
In these cases, the system responds to the slowly varying conditions with normal kinetic roughening
in a short time scale. This is expected for $|s(t)|\lesssim 1$.

When $|s(t)|\ll 1$, the roughness evolves similarly to the constant temperature case. This is
the case of small $\frac{E}{k_BT}$ (very high temperatures, in which the surface is always very smooth)
and of very slow changes of temperature (when changes in $\ln{T}$ are much smaller than $1$ after
growth of several layers).

If $|s(t)|\sim 1$, then the roughness changes are significantly different from the constant temperature
case. If $s(t)$ is weakly dependent on time, then the rate of roughness increase is changed by a
nearly constant factor and $W$ will scale with the exponent $\beta$ of constant temperature.
On the other hand, if $s(t)$ varies in time, plots of $\log_{10}{W}\times\log_{10}{t}$ may
show effective growth exponents very different from $\beta$. If the temperature increases in time,
$s(t)$ is negative and the roughness may decrease in time.

Finally, if $|s(t)|\gg 1$, the growth conditions are rapidly changing
and Eq. (\ref{dwdt}) suggests that it controls the roughness evolution. Under these conditions,
it is not expected that the extension of Eq. (\ref{growthcdlm}) to time varying $R$ is correct, since the
system may not be able to adapt to varying growth conditions with its normal kinetic roughnening
behavior. Thus, Eq. (\ref{dwdt}) may not apply. Moreover, the hypothesis of rapidly changing conditions
may also rule out the thermal equilibrium hypothesis of Eq. (\ref{defD}) adopted along the whole
surface. For this reason, the case  $|s(t)|\gg 1$ is not considered here.

\section{Deposition with decreasing temperature}
\label{decreasingT}

These are situations with slow down of the diffusion process, thus roughening is facilitated as
the deposition evolves.

\subsection{Temperature variation}
\label{tempvariation}

First we consider linearly decreasing temperature during the time interval for deposition
of $t_{max}={10}^4 ML$, numbered cases (I) and (II), whose conditions are specified in Table 1.
This value of $t_{max}$ is reasonable for mesoscopic films, giving thicknesses of a few micrometers
for most materials. Figs. 3a shows the time evolution of the parameter $R$ and of the temperature in
those cases.

\begin{figure}
\begin{center}
\includegraphics [width=7.5cm] {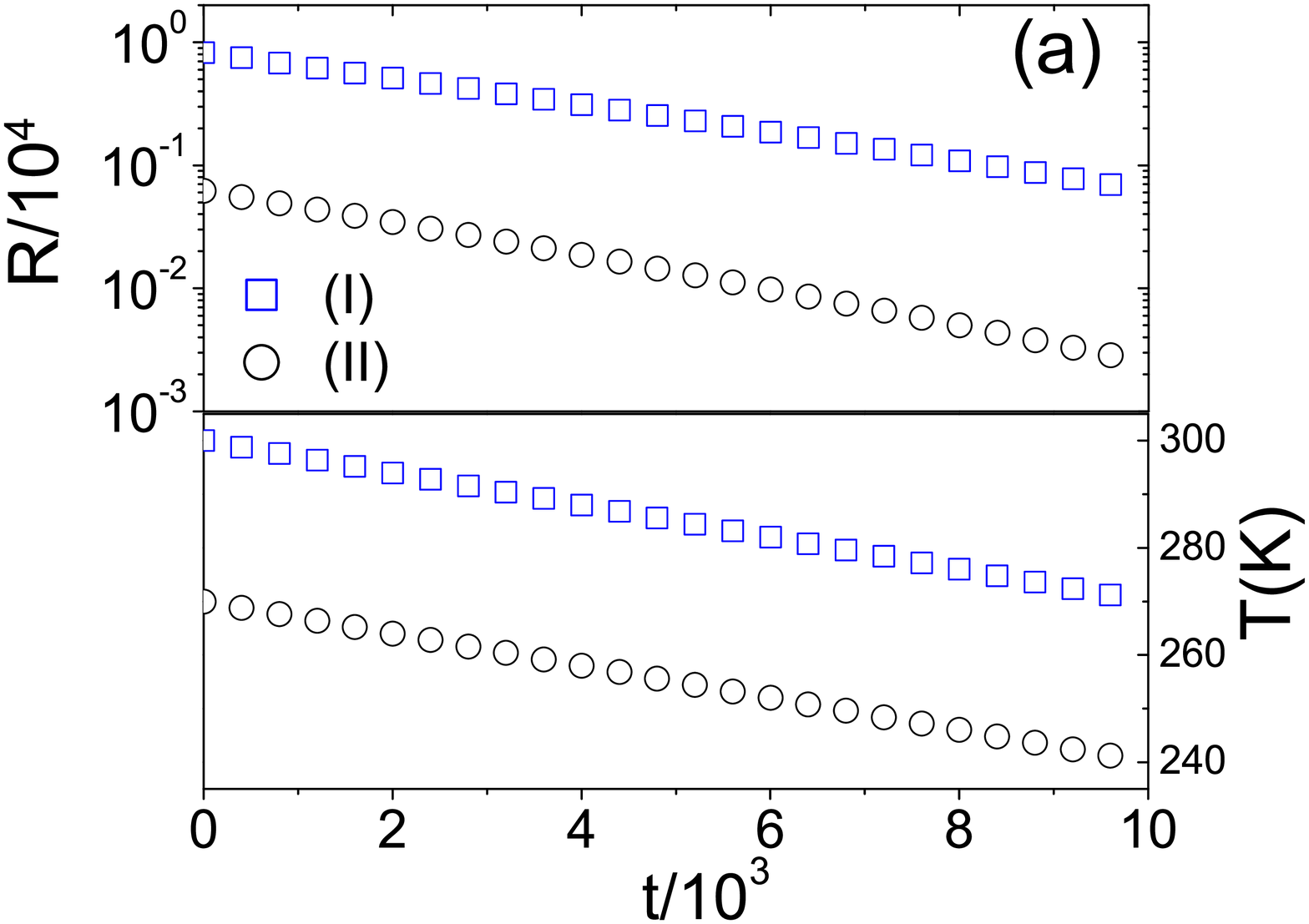}
\includegraphics [width=7.5cm] {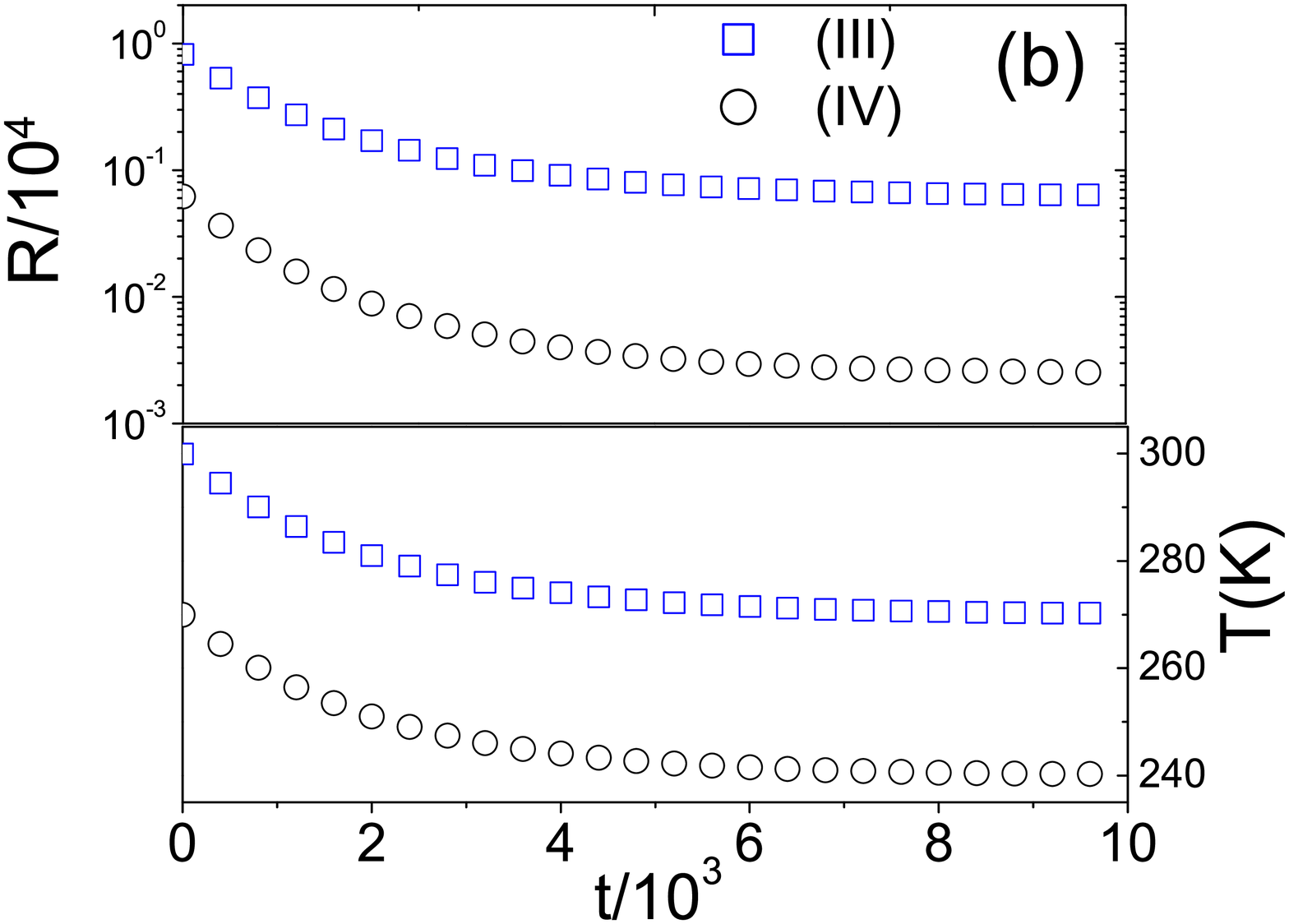}
\caption{(a) Parameter $R$ as a function of the thickness, considering $E=0.6$eV, $\nu_{0}/F$ = $10^{14}$,
and temperature variation of cases (I) and (II). Squares correspond to case (I) and circles to case (II).
(b) Parameter $R$ as a function of the thickness, considering $E=0.6$eV, $\nu_{0}/F$ = $10^{14}$,
and temperature variation of cases (III) and (IV). Squares correspond to case (III) and circles to case (IV).}
\label{fig3}
\end{center}
\end{figure}

We also consider cases of exponential convergence of the temperature to the final value,
with a characteristic time/thickness $t_c = 2000 ML$, so that:
\begin{equation}
T=T_F+\left( T_I-T_F\right)\exp{\left( -t/t_c\right)} .
\label{tempexp}
\end{equation}
The conditions of cases (III) and (IV) considered here are presented in Table 1.
Deposition of $t_{max}={10}^4 ML=5t_c$ layers is considered, thus the final temperatures of
the films are very close to $T_F$. These cases represent systems exchanging heat by
conduction with a colder reservoir (e. g. an initially heated system separated from the
surroundings by a conducting wall).

\begin{table} 

 \centering

 \begin{tabular}{|c|c|c|c|c|} 

 \hline 

 Case Number & Temperature Variation & Thickness Dependence & $T_{I}$ (K) & $T_{F}$ (K) \\ 

 \hline
 \hline

 I & Decreasing & Linear & 300 & 270  \\ \hline
 II & Decreasing & Linear &270 & 240 \\ \hline
 III & Decreasing& Eq. \ref{tempexp} & 300 & 270 \\ \hline
 IV & Decreasing & Eq. \ref{tempexp} & 270 & 240  \\ \hline
 V & Increasing & Linear & 240 & 270 \\ \hline
 VI & Increasing & Linear & 270 & 300 \\ \hline
  VII & Increasing & Eq. \ref{tempexp}& 240 & 270 \\ \hline
 VIII & Increasing & Eq. \ref{tempexp} & 270 & 300 \\ \hline
 \end{tabular}
 
 \caption{In all the cases we considered the deposition of $t_{max}={10}^4 ML$.}
 \label{tab}
 \end{table}

Fig. 3b shows the corresponding time evolution of the parameter $R$ and temperature.
The main difference from the linear decays of cases (I) and (II) is that $R$ rapidly decreases at short times
(up to $t\sim t_c$) and slowly decreases during most of the deposition time. When $t\sim t_{max}$,
the temperature is approximately constant ($T_F$).

In cases (I) to (IV), $E=0.6 eV$ in the room temperature range, thus
$E/k_BT\sim 23$. Maximal temperature changes are close to $10\%$ of the initial ones, from $1$
to ${10}^4$ monolayers, thus $\frac{d\ln{T}}{d\ln{t}} \sim 0.01$. This gives $s\sim 0.2$ [Eq. (\ref{dwdt})],
which is expected to give a significant change in the roughness scaling (Sec. \ref{theoretical}).
Depending on the particular form that the temperature varies, larger or smaller values of $s$
may be obtained in different time ranges, as shown below.

The control of these temperature changes are realistic for thin film growth. For comparision, in film deposition
with the substrate subject to a temperature gradient, much larger temperature differences are
stablished \cite{schwickert,candia}. An example is $FePt$ film growth with the temperature varying
from $250 {}^oC$ to $600 {}^oC$ along a substrate distance smaller than $1cm$ \cite{schwickert}).

\subsection{Global roughness scaling}
\label{globaltempdecrease}

Fig. 4a shows the evolution of the global roughness of the films in cases (I) and (II)
and the theoretical predictions from Eq. (\ref{growthcdlm}) with the time-dependent $R$ shown in Fig. 3a.
Those predictions are close to the simulation data when the roughness is larger than $1$.
The discrepancy for $W\lesssim 1$ is expected because scaling relations are not expected to apply for
small roughness. The local slope of the plots in Fig. 4a are effective exponents $\beta_{eff}$.

\begin{figure}
\begin{center}
\includegraphics [width=7.5cm] {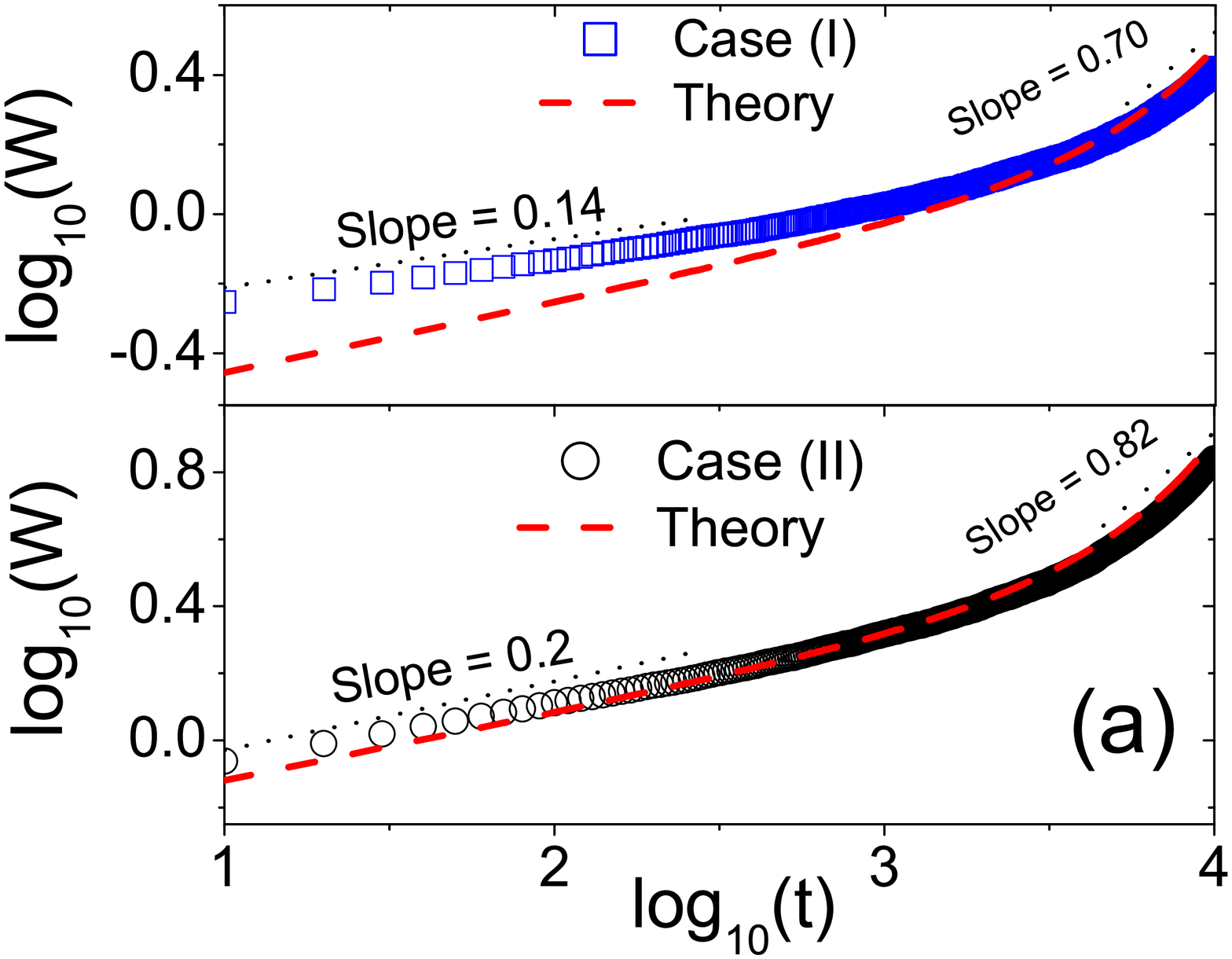}
\includegraphics [width=7.5cm] {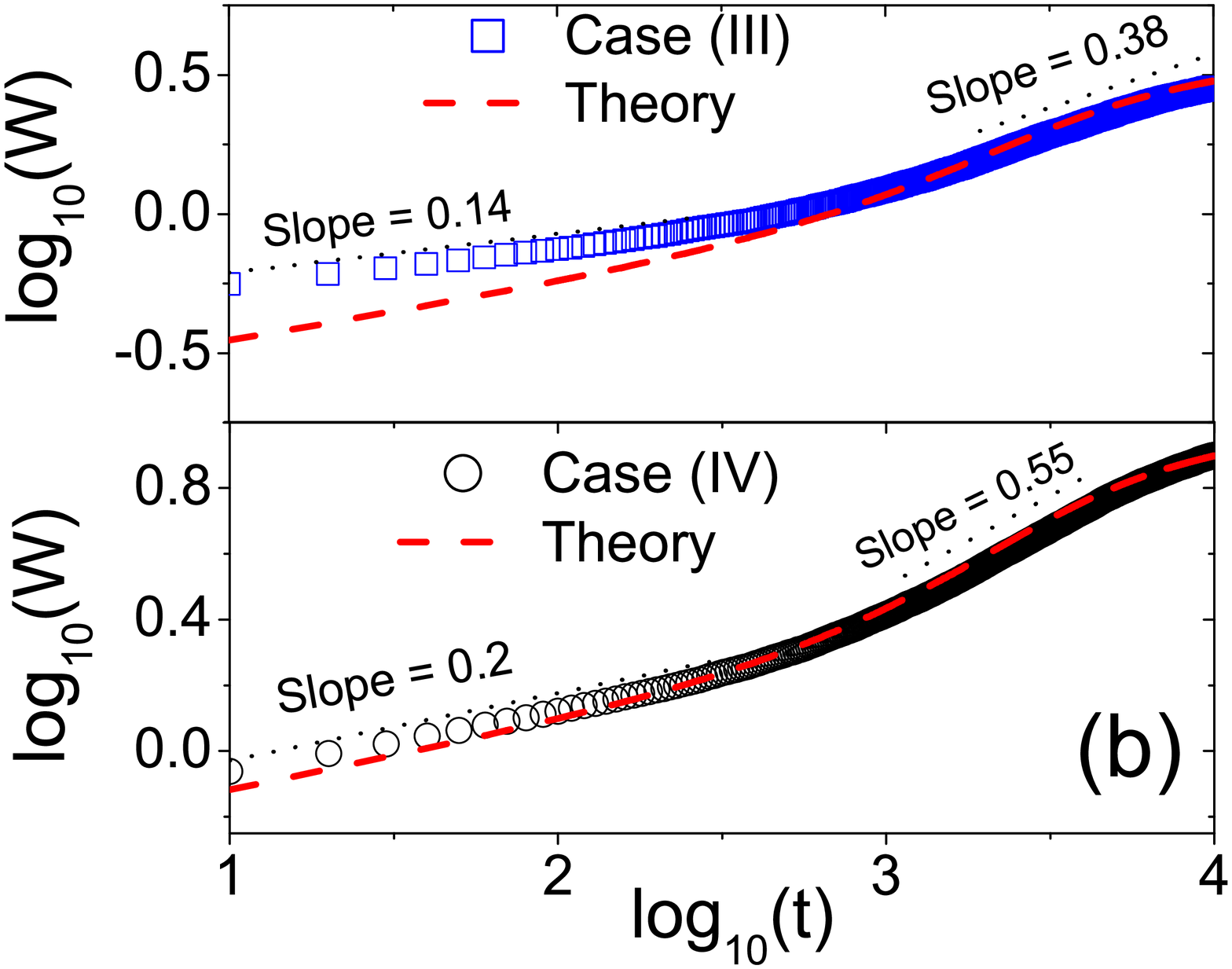}
\caption{(a) Global roughness as a function of the thickness for cases (I) and (II).
Squares correspond to case (I) and circles to case (II). The dashed line
indicate the theoretical predictions from Eq. (\ref{growthcdlm}) with the time-dependent $R$ shown in Fig. 3a.
(b) Global roughness as a function of the thickness for cases (III) and (IV).
Squares correspond to case (III) and circles to case (IV). The dashed line
indicate the theoretical predictions from Eq. (\ref{growthcdlm}) with the time-dependent $R$ shown in Fig. 3b.}
\label{fig4}
\end{center}
\end{figure}

For small thicknesses ($t\leq {10}^3 ML$), the slope $\beta_{eff}$ is close to the
VLDS value $\beta\approx 0.2$. This corresponds to one tenth of the total
growth time and the temperature change is only $3K$, thus it is a growth with nearly constant temperature.
For case (I), the roughness is typically smaller than unity, thus it is not a true scaling region
where a reliable estimate of $\beta$ can be measured.
For larger thicknesses ($t\geq {10}^3 ML$), $W$ rapidly increases in cases (I) and (II).
For $5\times{10}^3 ML\leq t\leq {10}^4 ML$, the effective exponents are $\beta_{eff}\approx 0.70$ and
$\beta_{eff}\approx 0.82$, respectively (see Fig. 4a).

The evolution of the global roughness for cases (III) and (IV) is shown in Fig. 4b and
also compared with the extension of Eq. (\ref{growthcdlm}) to time-dependent $R$ (Fig. 3b).
Again, $W$ increases slowly for small thicknesses, but shows large effective exponents $\beta_{eff}$
for thicknesses above ${10}^3 ML$. In case (IV), where $R$ decreases almost two orders of magnitude
during the film growth, $\beta_{eff}>1/2$ is found, similarly to cases (I) and (II).
As $t$ approaches  $t_{max}$, the slopes of the plots in Fig. 4b tend to decrease due to the temperature
saturation.

The continuously increasing $\beta_{eff}$, which attains large values at longer times, represents
a nontrivial evolution of the roughness. This is solely a consequence of the time decreasing
temperature during deposition, which reduces diffusion lengths and facilitates roughening.
The agreement with the theoretical approach of Sec. \ref{theoretical} supports this interpretation.

Some models and experiments show large values of $\beta_{eff}$, but for different reasons. An example is
Ag/Ag(100) growth, where mound steepening leads to a crossover from very smooth surfaces (up to $25 ML$)
to very rough ones (up to $1000 ML$) at $300K$ \cite{caspersen}. In that case, the presence
of ES barriers \cite{es} is responsible for the crossover. Some systems also have $\beta > 1/2$ as a consequence
of  preferential aggregation at surface hills instead of valleys (possibly with surface instability) or
shadowing effects \cite{yanguas,Woo}. However, in our model, none of those mechanisms are present.


\begin{figure}
\includegraphics [width=8.5cm] {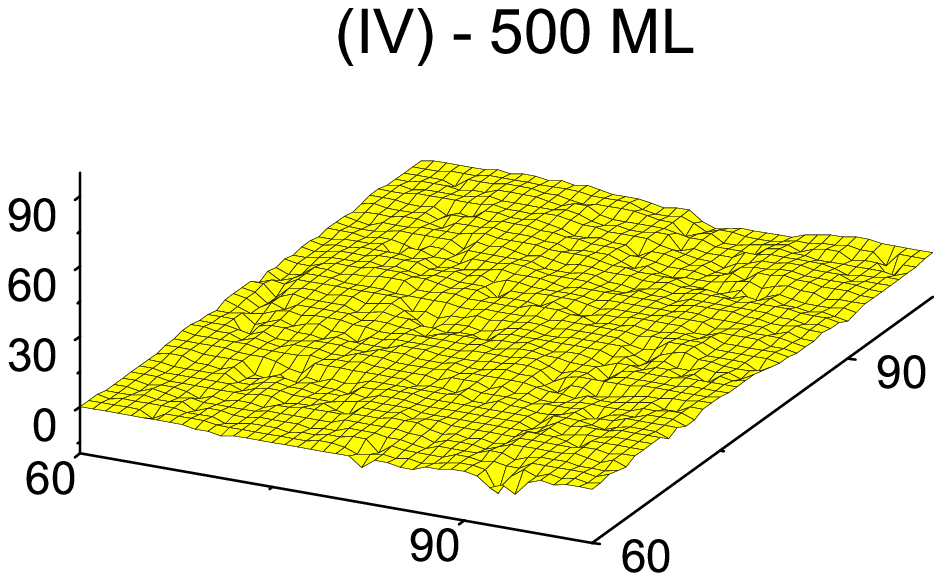}
\includegraphics [width=8.5cm] {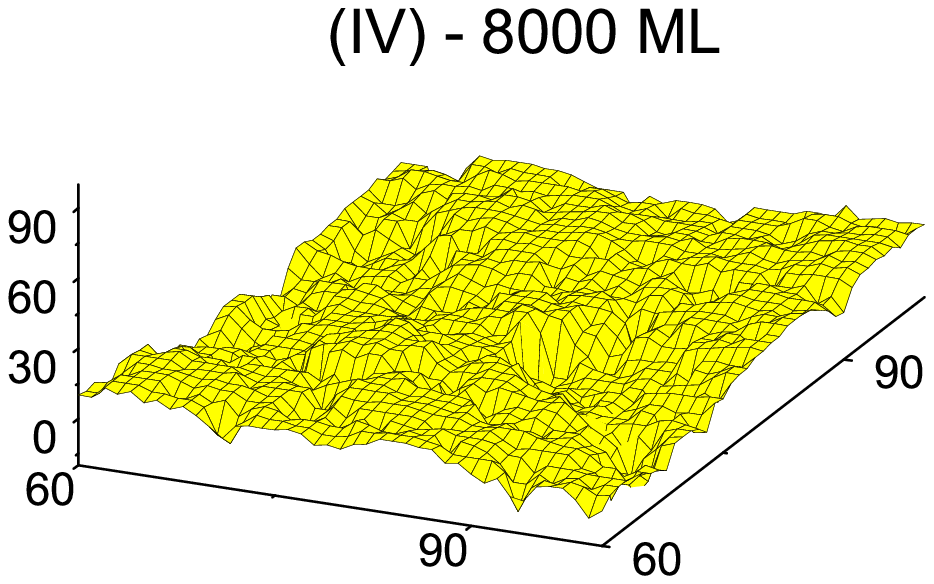}
\includegraphics [width=8.5cm] {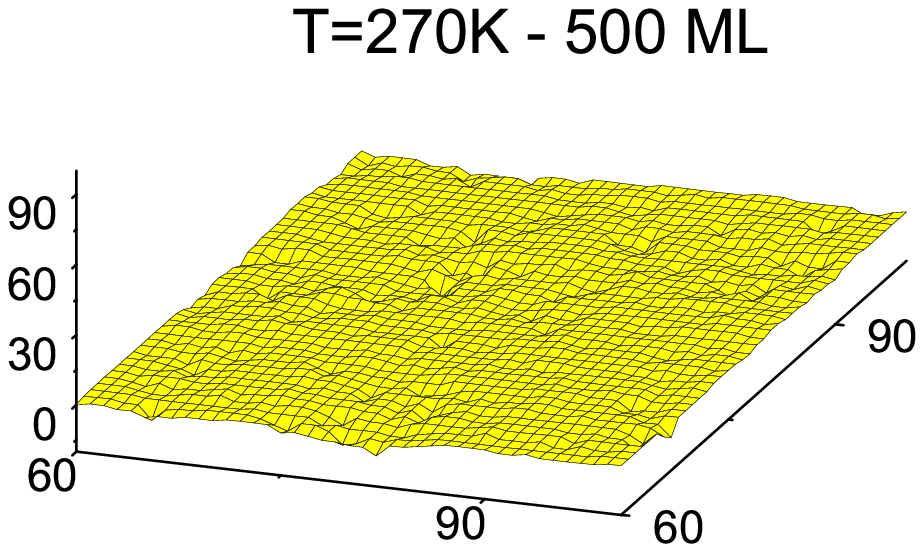}
\includegraphics [width=8.5cm] {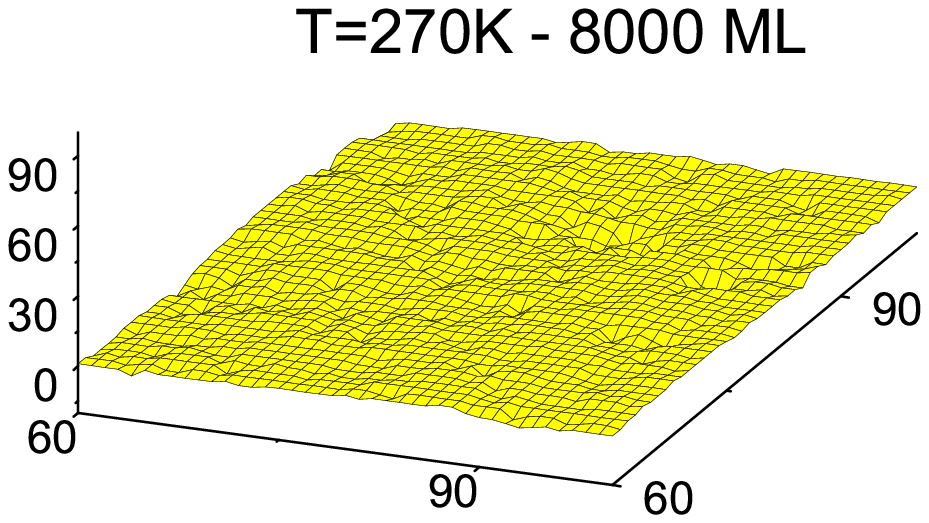}
\caption{Surface topography of films with thicknesses $t=500$ML and $t=8000$ML in (top) case (IV) and (bottom) with fixed temperature $T=270$K.}
\label{fig5}
\end{figure}
Fig. 5 shows the surface topography of films with thicknesses $t=500 ML$ and
$t=8000 ML$ in case (IV) and with fixed temperature $T=270 K$.
The film grown in fixed temperature is always smooth, but
a remarkable roughening is observed in the film grown with decreasing temperature.
In the latter, there is no evidence of mound formation, columnar growth, or any other geometrical
feature that frequently explains the large values of $\beta_{eff}$. Instead,
the nontrivial evolution of the roughness is a consequence of the decreasing temperature.

We also recall that the scaling in systems with competitive growth dynamics is very different
from Figs. 4a-b, since they usually show two scaling regions with $\beta_{eff}\leq 1/2$.
Moreover, large $\beta$ is not necessarily related to large roughness. For instance,
very rough deposits are produced by grain deposition models \cite{grains} and have $\beta\sim 0.1$, which is
much smaller than the value of the asymptotic Kardar-Parisi-Zhang class \cite{kpz}.

For the above reasons, if a nontrivial increase of the roughness is observed in an experimental work,
one must consider the possibility that the temperature is changing during the deposition, since this seems
to be the simplest mechanism responsible for that feature. Alternatively,
the possibility of time-increasing adsorption rates may be considered, as in Ref. \protect\cite{PB}
(where $\beta \approx 0.5$ was obtained), since they also lead to time-decreasing $R$.

\subsection{Local roughness scaling}
\label{localtempdecrease}

The local roughness in cases (I) and (II) is shown in Fig. 6a as a function of the box size $r$,
for several thicknesses in the range $500 ML<t<8000 ML$.
There is a significant split of the curves for small $r$, in contrast
to the small split shown in Fig. 2 for constant temperature (note that vertical and horizontal scales are
the same in Figs. 2 and 6a).
This is an AS feature, with the FV relation [Eq. (\ref{fvlocal})] having time-dependent amplitude $A$.

\begin{figure}
\begin{center}
\includegraphics [width=7.5cm] {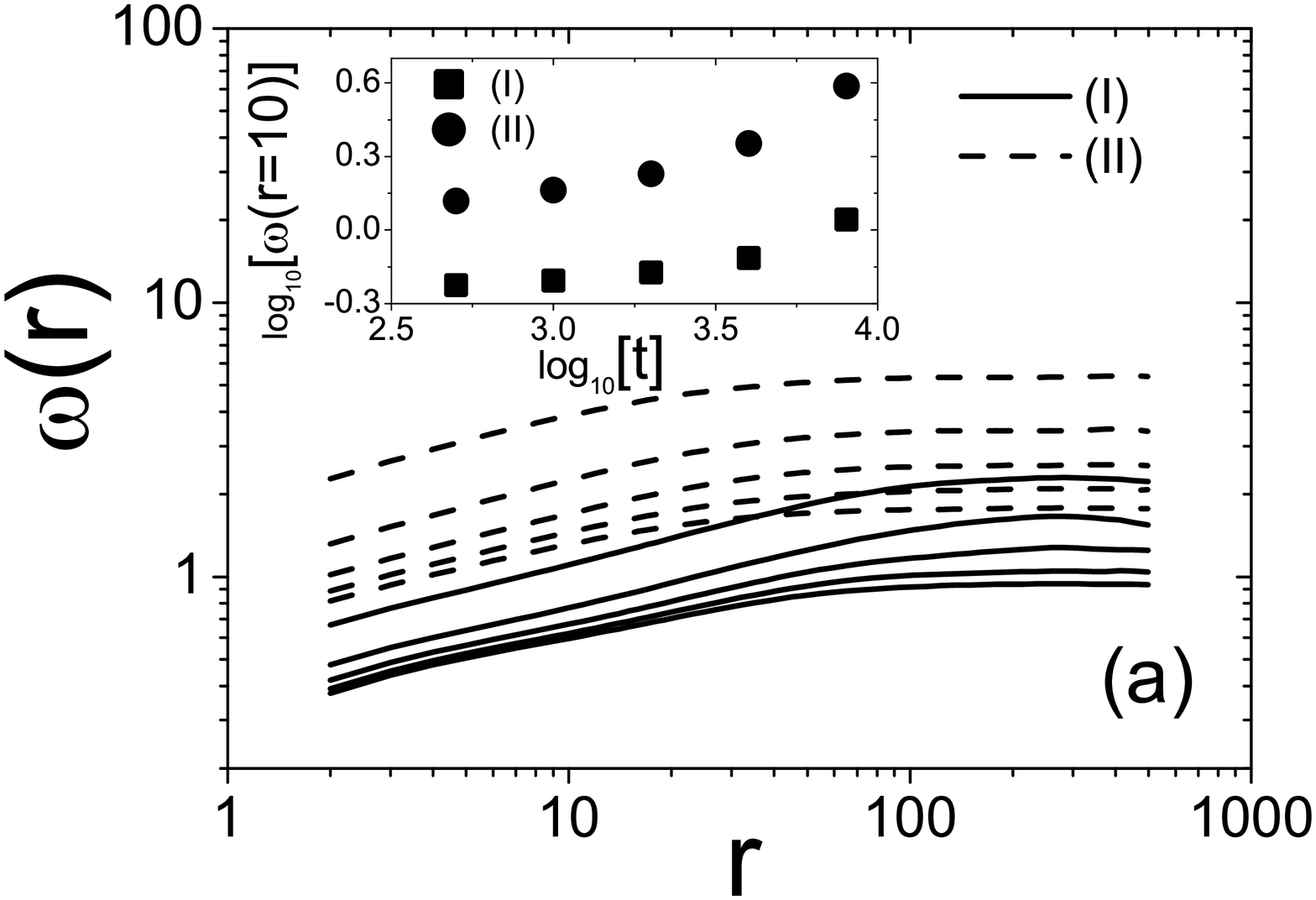}
\includegraphics [width=7.5cm] {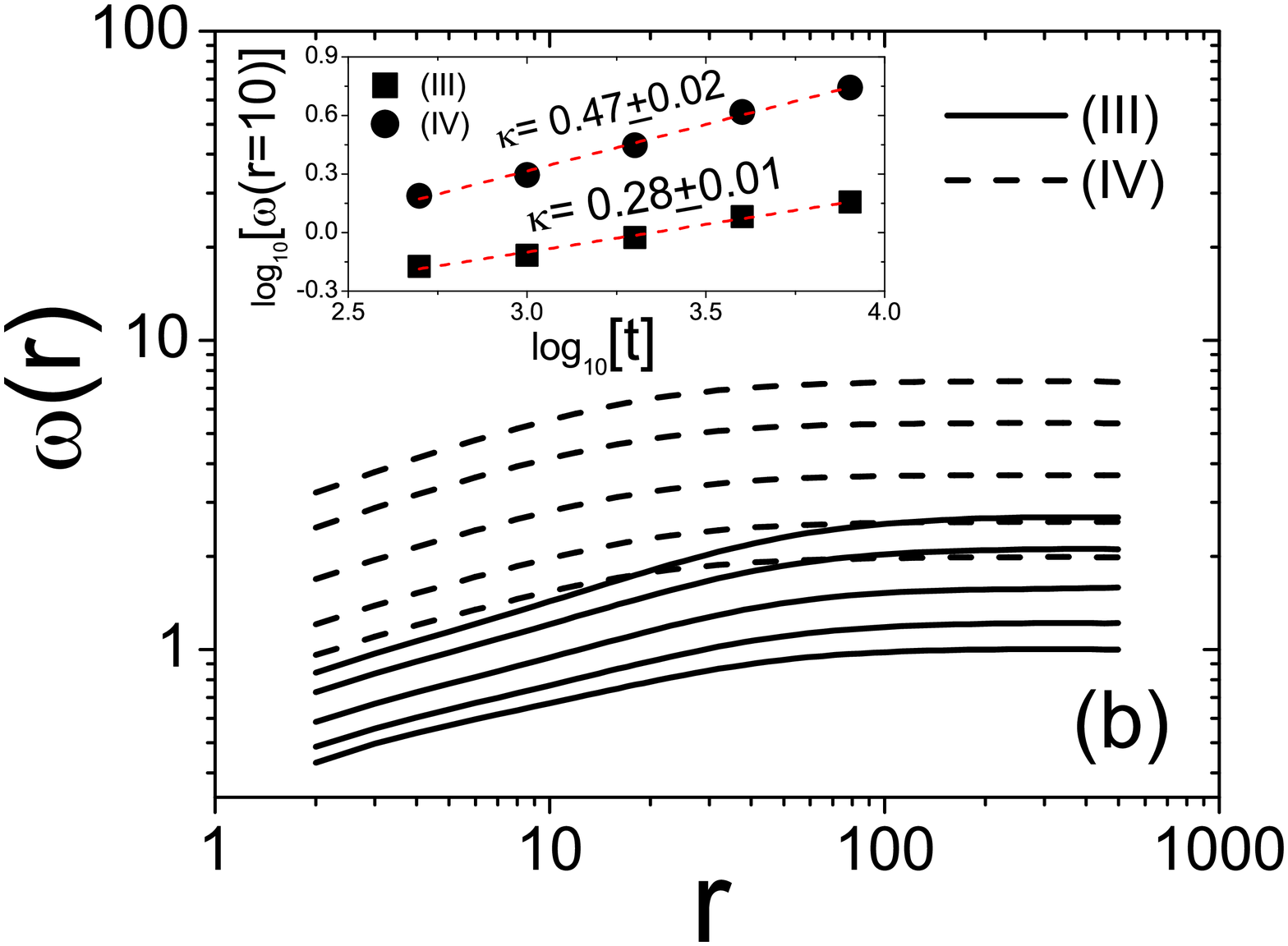}
\caption{(a) Local roughness as a function of box size for cases (I) (solid lines) and (II) (dashed lines)
at times $t=500$ML, $1000$ML, $2000$ML, $4000$ML and $8000$ML, from bottom to top.
The inset shows $\log_{10}{(\omega)}$ as a function of $\log_{10}{(t)}$ for $r=10$ in cases (I) (squares)
and (II) (circles).
(b) Local roughness as a function of box size for cases (III) and (IV), respectively,
at times $t=500$ML, $1000$ML, $2000$ML, $4000$ML and $8000$ML, from bottom to top.
The inset shows $\log_{10}{(\omega)}$ as a function of $\log_{10}{(t)}$ for $r=10$ in cases (III) (squares)
and (IV) (circles). The dashed lines indicate the linear fits for the $\kappa$ evaluation.
The corresponding values of $\kappa$ are also shown.}
\end{center}
\label{fig6}
\end{figure}

Fig. 6b shows the local roughness as a function of box size for cases (III) and (IV),
respectively. These plots also show evidence of AS due to the split of the curves for small $r$.
This trend persists for thicknesses $t > 2t_c$, where the temperature is changing very slowly.

The AS, corresponding to the continuous increase of local slopes, is another feature intrinsically
related to the slow down of surface diffusion. Indeed, reduced adatom mobility is known to facilitate
the formation of steep surface features at small lengthscales.

For small and constant $r$, the time scaling of the amplitude $A$ [Eqs. (\ref{fvlocal}) and (\ref{defkappa})]
characterizes the AS. The inset of Fig. 6a shows $w(r=10,t)$ (proportional to $A$) as a function of time
in cases (I) and (II). It increases faster than the power-law of Eq. (\ref{defkappa}),
particularly for the longest times. On the other hand, the inset of Fig. 6b shows the scaling of $w(r=10,t)$ for
cases (III) and (IV), which approximately follow power laws with
anomaly exponents $\kappa =0.28\pm 0.01$ (III) and $\kappa=0.47\pm 0.02$ (IV).
This reinforces the evidence of AS in plausible conditions for experimental work only due to
the decreasing temperature.

It is important to stress that the AS found in cases (III) and (IV) is an apparent scaling
restricted to the thickness range studied here. At long times, the temperature saturates at $T_F$,
thus the deposit will attain the characteristic features of films grown with fixed temperature.
Such films do not have AS, but have normal VLDS scaling (Sec. \ref{local}). Furthermore, in cases (I) and (II),
the temperature decrease is necessarily limited to a finite time range, thus there is no
asymptotic scaling. These features contrast with the true AS of the lattice models in
Ref. \protect\cite{anomcompet}, but those models were not so closely related to real film growth.

Finally, we observe that the local roughness exponents measured in Figs. 6a and 6b
are close to $1/3$, similarly to the constant temperature case (Sec. \ref{local}).

\section{Deposition with increasing temperature}
\label{increasingT}

We also consider two cases of linearly increasing temperature, numbered (V) and (VI) in Table 1,
again from $t=0$ to  $t_{max}={10}^4 ML$.
Fig. 7a shows the time evolution of the parameter $R$ and of the temperature.
The situation with exponential convergence of temperature [Eq. (\ref{tempexp}) with $T_I<T_F$]
is considered in cases (VII) and (VIII), also described in Table 1,
with characteristic decay time $t_c = 2000 ML$.
Fig. 7b shows the corresponding time evolution of the parameter $R$ and of the temperature.

\begin{figure}
\begin{center}
\includegraphics [width=7.5cm] {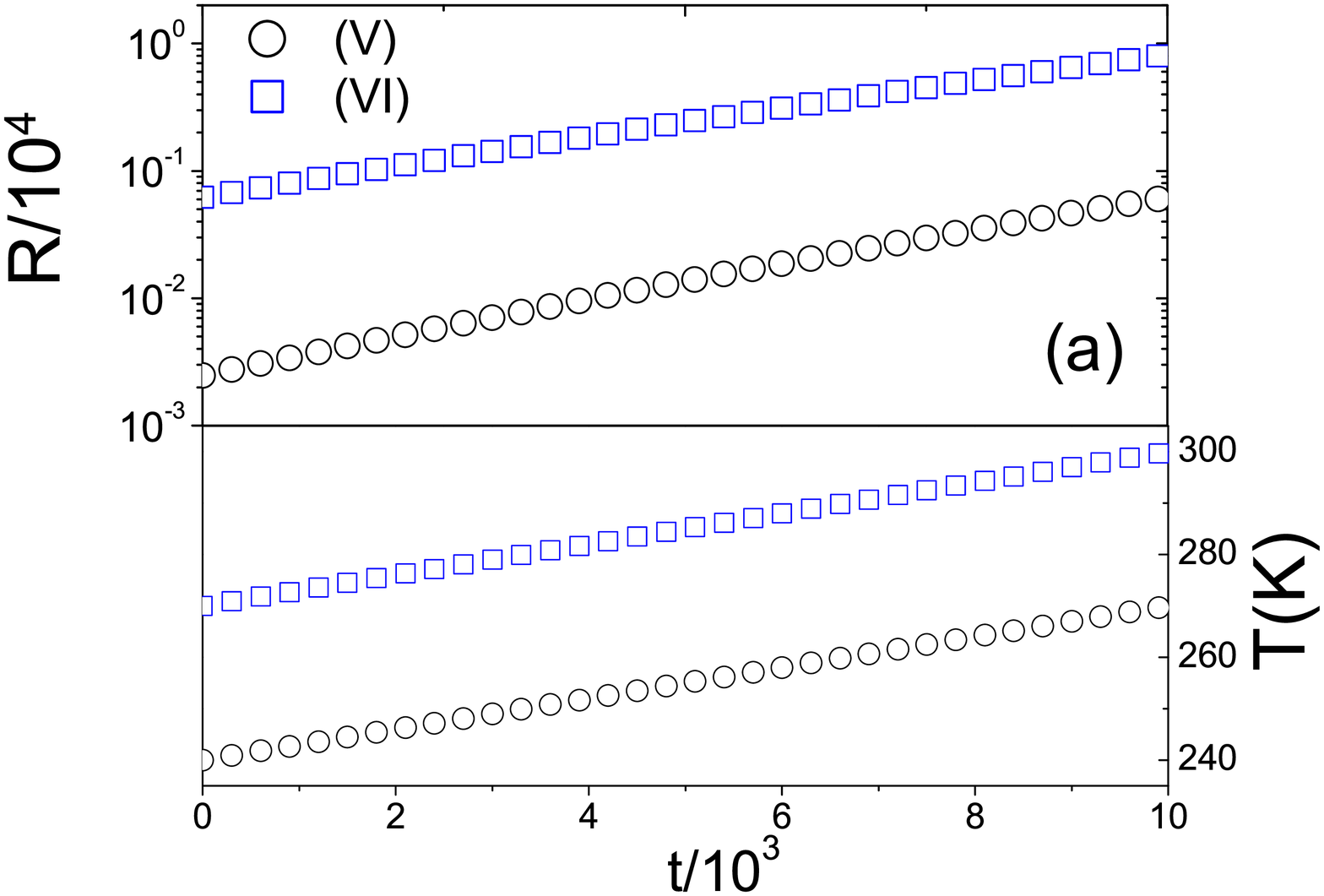}
\includegraphics [width=7.5cm] {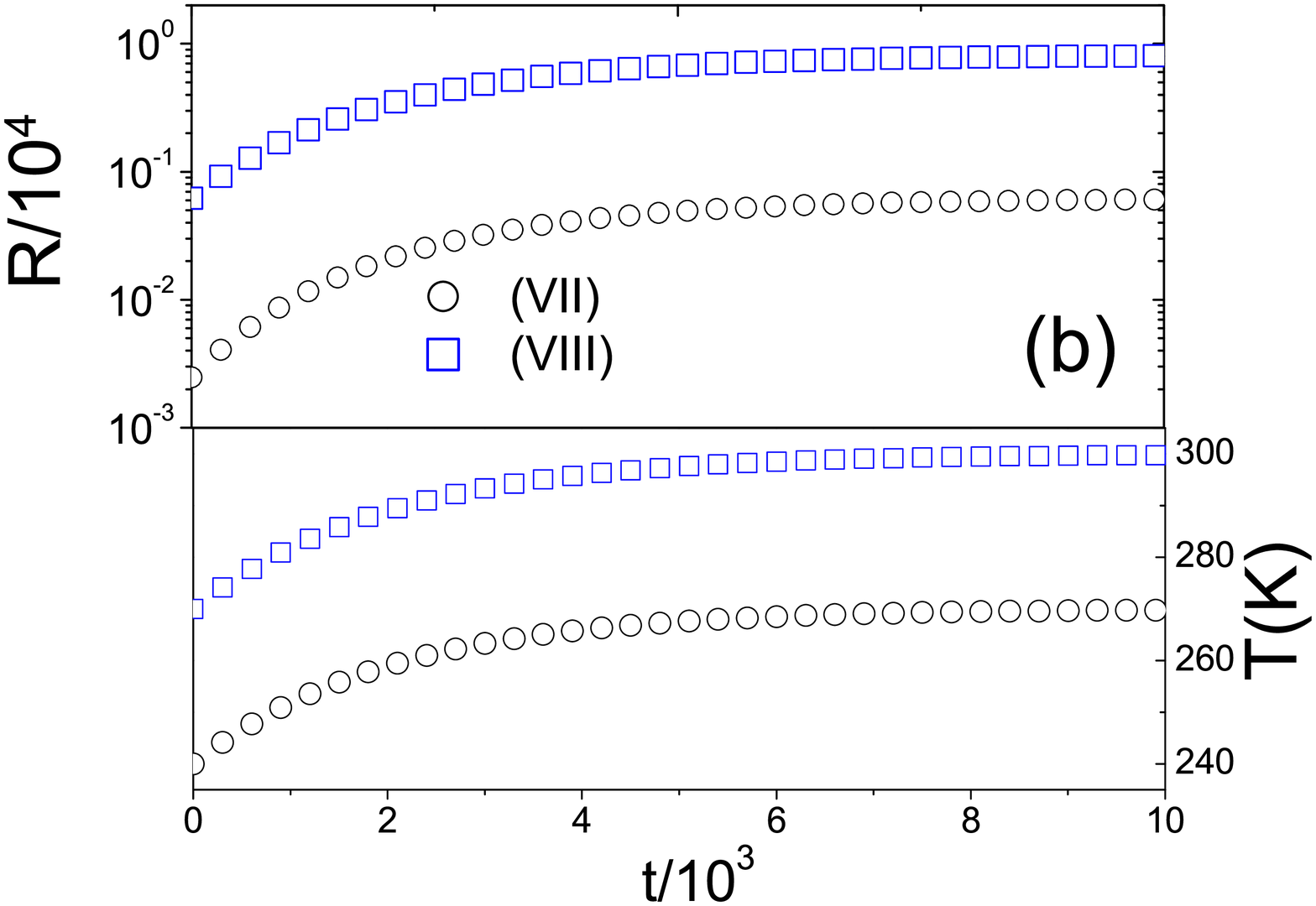}
\caption{(a) Parameter $R$ as a function of the thickness, considering $E=0.6$eV, $\nu_{0}/F$ = $10^{14}$,
and temperature variation of cases (V) and (VI). Circles correspond to case (V) and squares to case (VI).
(b) Parameter $R$ as a function of the thickness, considering $E=0.6$eV, $\nu_{0}/F$ = $10^{14}$,
and temperature variation of cases (VII) and (VIII). Circles correspond to case (VII) and squares to case (VIII)}
\label{fig7}
\end{center}
\end{figure}
Fig. 8a shows the evolution of the global roughness in cases (V) and (VI) and
the comparison with Eq. (\ref{growthcdlm}) extended to time-dependent $R$ case shown in Fig. 7a.
Again, deviations from the theoretical prediction are large only when the global roughness
is very small ($W\lesssim 1$, i. e. smaller than the lattice parameter).

The nontrivial result in Fig. 8a is the nonmonotonic variation of the roughness.
It increases for short times due to spread of correlations and excitation of modes of increasing
wavelength, as usual in kinetic roughening \cite{barabasi,nt}. However, it attains a maximum and
begins to decrease due to the smoothing effect of large diffusion coefficients. The effect
is more pronounced for smaller temperatures because the relative increase of $R$ is larger.

\begin{figure}
\begin{center}
\includegraphics [width=7.5cm] {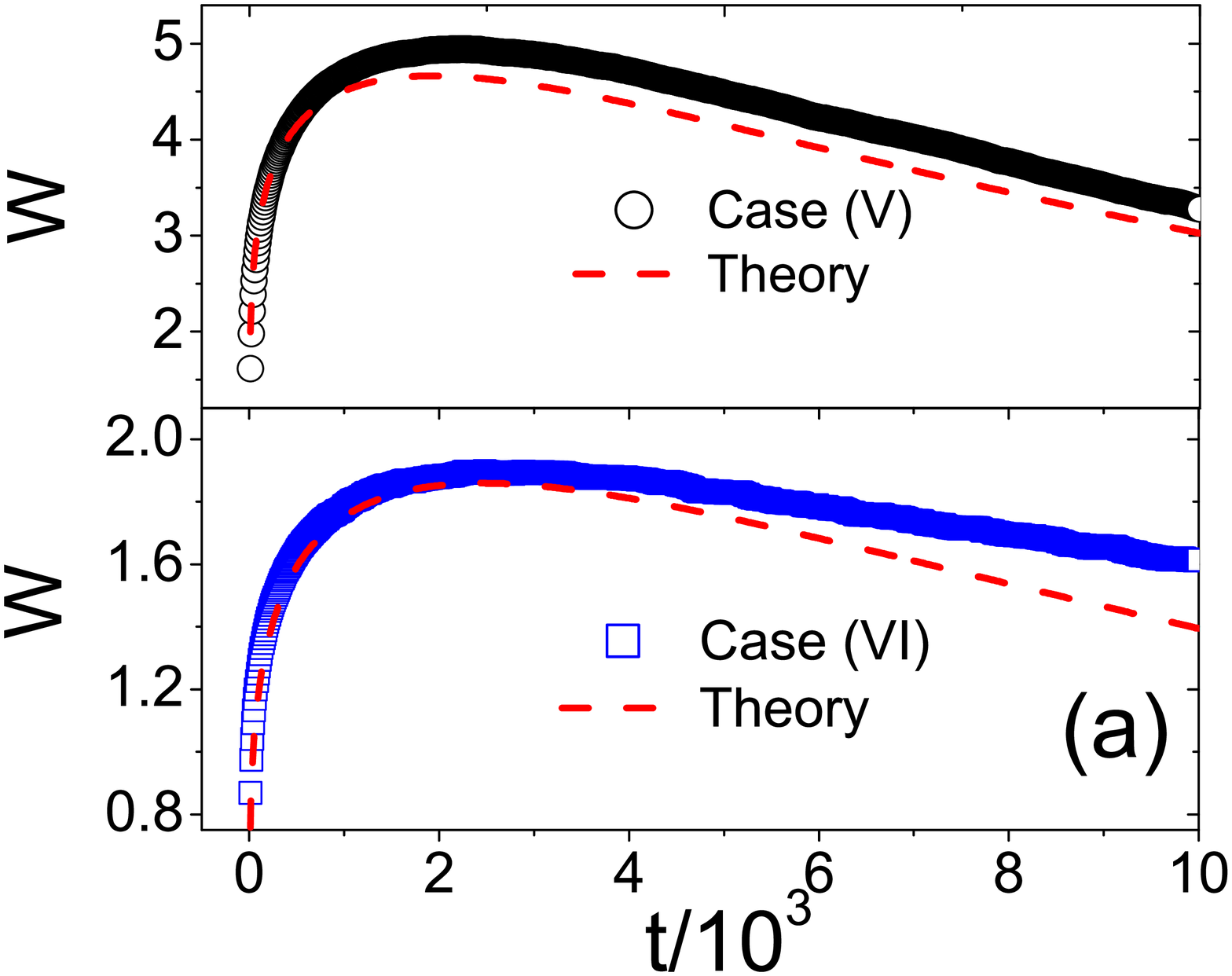}
\includegraphics [width=7.5cm] {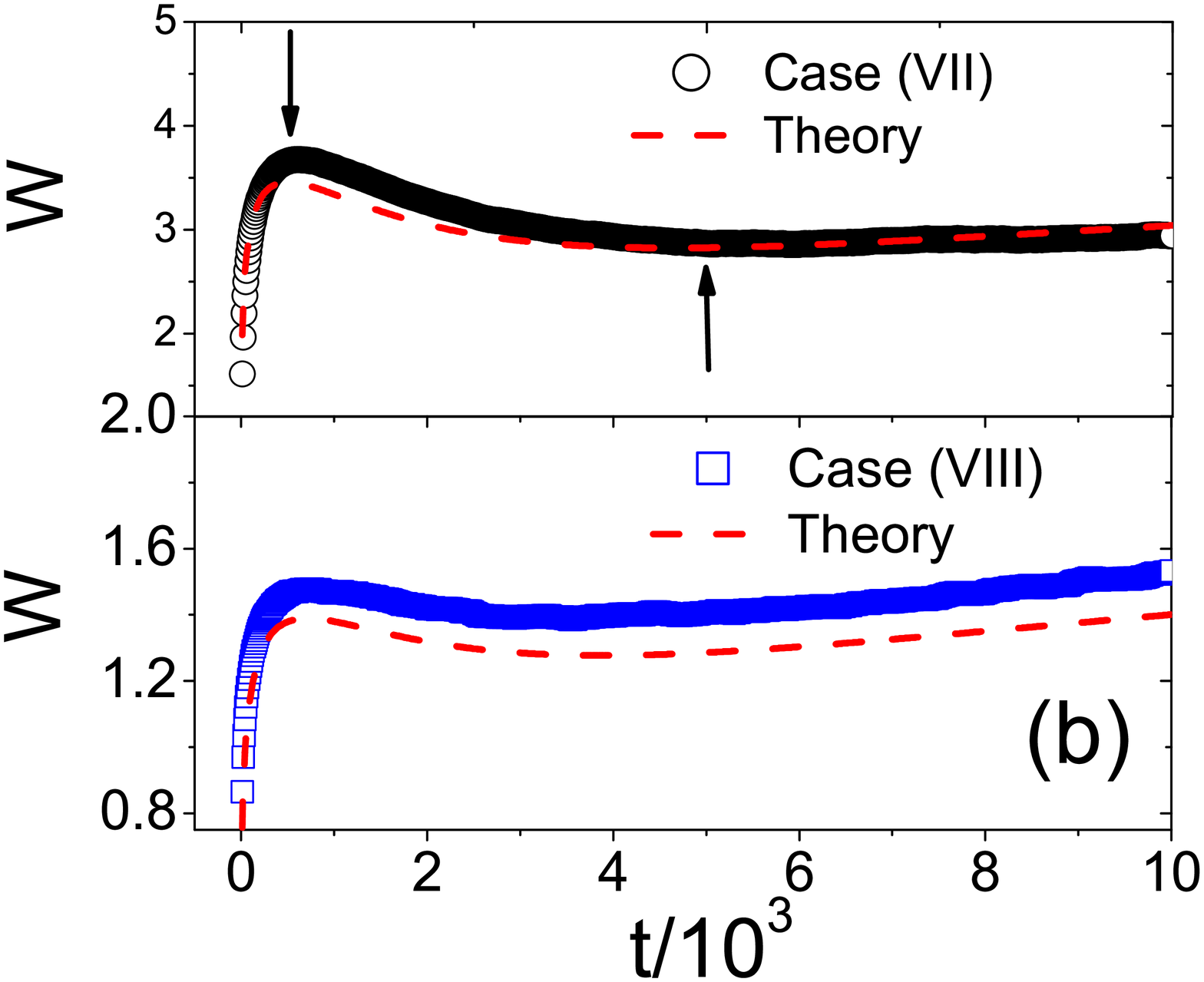}
\caption{(a) Global roughness as a function of the thickness for cases (V) and (VI).
Circles correspond to case (V) and squares to case (VI). The dashed line
indicate the theoretical predictions from Eq. (\ref{growthcdlm}) with the time-dependent $R$ shown in Fig. 7a.
(b) Global roughness as a function of the thickness for cases (VII) and (VIII). The dashed line
indicate the theoretical predictions from Eq. (\ref{growthcdlm}) with the time-dependent $R$ shown in Fig. 7b.
The arrows indicate a maximum and a minimum
of global roughness. Circles correspond to case (VII) and squares to case (VIII).}
\label{fig8}
\end{center}
\end{figure}
Fig. 8b shows the evolution of the global roughness in cases (VII) and (VIII),
with the comparison with the theoretical prediction for time-dependent $R$ case shown in Fig. 7b.
$W$ presents a maximum at small thicknesses in the former and a plateau in the latter. These effects are
similar to those of the linearly increasing $T$ [(V) and (VI)]. However, for larger thicknesses,
the roughness slowly increases. This is expected because
the temperature saturates at $t\sim t_{max}$, thus the usual feature of kinetic roughening
(roughness increasing in time) is observed.

The presence of extremal values in the roughness (a maximum and a minimum) was already
shown in other models, but not as an effect of changing temperature. For instance,
in an electrodeposition model of Ref. \protect\cite{santos2002}, the adsorption rate had a maximum,
which corresponds to a minimum of an effective diffusion-to-deposition
rate $R$. The increase of $R$ after that minimum seems to be responsible for
the effect of reducing the roughness in a certain time window, similarly to our model.
Pal and Landau \cite{pal,landau} also showed that deposition with large $R$ may lead to roughness
oscillations at short times, due to fluctuations between partial and complete filling of the first atomic
layers. However, those oscillations disappear after deposition of approximately $10$ layers (maybe less),
in contrast to our models.

\section{Possible extensions of the model}
\label{extensions}

As shown in Secs. \ref{decreasingT} and \ref{increasingT}, the theoretical predictions of Sec.
\ref{theoretical} for time-dependent $R$ fit the simulation data quite well when the roughness
is not very small. Thus, that approach may help to predict conditions for observing the same
phenomena observed here, such as AS and rapid roughness increase (large $\beta_{eff}$)
for decreasing temperature and roughness oscillations for increasing temperature.

In a more complex growth model, other temperature-dependent parameters are relevant, typically in
the form of Boltzmann factors involving other activation energies. Let us recall, for instance,
the studies of submonolayer island dynamics where all adatoms are mobile. This is the case of
the Clarke-Vvedensky model \cite{cv}, in which activation energies are proportional to the
number of lateral neighbors. The additional parameter $\epsilon$ is related to the lateral binding energy
and the scaling of average island size contains factors in the form $R^{a}\epsilon^{b}$,
with exponents $a$ and $b$ of order $1$, related to the nature of the model and possibly
to the lattice structure \cite{bartelt1995,submonorev}.

In the growth of a thin film, assuming that an additional parameter $\epsilon$ is present,
the FV relation for short times and constant temperature is expected to give
\begin{equation}
W = C \frac{t^{\beta}}{R^{\delta}\epsilon^{\gamma}} ,
\label{fvother}
\end{equation}
where $\beta$ is the exponent of the universality class of the model (possibly not VLDS) and
$\delta$ and $\gamma$ are exponents of order $1$.

The brackets in the right-hand side of Eq. (\ref{dwdt}) separate the roughness variation in two terms,
the second one related to variable temperature. Using Eq. (\ref{fvother}), this time-dependent term is
\begin{equation}
s (t)\equiv \left(\frac{E}{k_BT}-\gamma /\delta \ln{\epsilon}\right)\frac{d\ln{T}}{d\ln{t}} .
\label{rother}
\end{equation}
Again, $s (t)$ is given by the product of energy-temperature ratios
($E/\left( k_BT\right)$, $\ln{\epsilon}$) and the temperature evolution expressed as $\frac{d\ln{T}}{d\ln{t}}$.
The conditions in which the temperature variation significantly affects the roughness are similar
to our model: $|s (t)|\sim 1$. Thus, in room temperature conditions and with activation energy sets similar to
the one studied here, the same features of the roughness evolution are expected.

These results suggest the investigation of temperature variation during the deposition of a film
in all cases where surface dynamics is dominated by adatom diffusion and where features such as AS,
large growth exponents (possibly in a small time range), or nonmonotonic variation of roughness are observed.

\section{Conclusion}
\label{conclusion}

We studied a thin film growth model in which the surface diffusion coefficients of adsorbed
species are related to the substrate/film temperature in conditions where that temperature
increases or decreases during the deposition. The model assumes that only adatoms in terraces can move,
so that a single parameter determines scaling properties.

Cases of temperature varying linearly
in time and of an exponential convergence to a final temperature were separately analyzed,
the later representing systems exchanging heat by conduction with a thermal reservoir.
In all cases, ${10}^4$ atomic layers were grown with temperatures in the
range of room temperature or below and variations during the growth up to $30 K$, which are
feasible conditions with many experimental techniques.

If the temperature decreases during the growth, the global roughness slowly increases at
short times but eventually turns to be a rapidly increasing function of time. In several cases,
effective growth exponents $\beta_{eff}>1/2$ are obtained in the thickness range ${10}^3 ML < t < {10}^4 ML$.
This shows that the simple physical mechanism of reducing the temperature during the growth
may lead to a nontrivial roughness evolution, which is usually related to more complicated mechanisms,
such as step energy barriers or shadowing.
The local roughness shows evidence of anomalous scaling in the thickness range
analyzed. In the cases of linear temperature decrease, the local roughness in a small box size may
increase in time even faster than a power law. In the cases of exponential convergence to a final
temperature, several values of the anomaly exponent may be obtained.

If the temperature increases during the growth, a non-monotonic evolution of the global roughness
may be observed. In cases of  linear temperature increase, a maximum of roughness is usually
observed, being more pronounced for lower temperature ranges. In the case of exponentially
converging temperature, a maximum and a minimum may be observed (eventually turning into a
plateau in the $log_{10}({W}) \times \log_{10}({t})$ plot). These features are interpreted as a consequence
of competition of kinetic roughening, which leads to roughness increase, and the
smoothing effect of increasing diffusion coefficients.

Stochastic growth equation approaches \cite{pradas} and lattice models \cite{anomcompet}
have already considered the effect of time-varying couplings in interface growth, with
emphasis on anomalous scaling properties. The main advance of the present work is to show
that a series of nontrivial features of surface roughness scaling (including anomalous
scaling) appear in realistic conditions of film growth. Moreover, our results shown very good agreement
with theoretical predictions in all the cases studied. In the
growth regime, that approach may help to predict conditions for observing the same
phenomena observed here, such as AS and rapid roughness increase (large $\beta_{eff}$)
for decreasing temperature and roughness oscillations for increasing temperature.

Finally, we note that results equivalent to the ones obtained with varying temperature
(i. e. varying $D$) may be obtained with variable flux $F$, since the main parameter of
our model is the ratio $R$. This is particularly interesting for possible experimental tests,
since controlling the deposition flux is usually easier than the temperature.

\section*{Acknowledgements}

The authors thank Prof. F. de B. Mota for fruitful discussions and valuable suggestions
to optimize the computational work.
The authors also acknowledge support from CNPq and FAPERJ (Brazilian agencies).


%

\end{document}